\documentclass[
    ,final            % use final for the camera ready runs
  ]
  {aipproc}

\layoutstyle{6x9}

\newcommand{\degr}{\hbox{$^\circ$}}

\begin{document}

\title{Astrophysical False Positives Encountered in Wide-Field Transit Searches}

\author{David Charbonneau}{
  address={California Institute of Technology, MC 105-24, 1200 E. California Blvd., Pasadena, CA 91125},
  ,email={dc@caltech.edu}}

\author{Timothy M. Brown}{
  address={High Altitude Observatory, Ntl Ctr for Atmospheric Research, P.O. Box 3000, Boulder, CO 80307},
  ,email={timbrown@hao.ucar.edu}}

\author{Edward W. Dunham}{
  address={Lowell Observatory, 1400 W. Mars Hill Road, Flagstaff, AZ 86001},
  ,email={gmand@lowell.edu}}

\author{David W. Latham}{
  address={Harvard-Smithsonian Ctr for Astrophysics, 60 Garden St., MS-20, Cambridge, MA 02138},
  ,email={dlatham@cfa.harvard.edu}}

\author{Dagny L. Looper}{
  address={California Institute of Technology, MC 105-24, 1200 E. California Blvd., Pasadena, CA 91125},
  ,email={dagny@caltech.edu}}

\author{Georgi Mandushev}{
  address={Lowell Observatory, 1400 W. Mars Hill Road, Flagstaff, AZ 86001},
  ,email={dunham@lowell.edu}}

\begin{abstract}
Wide-field photometric transit surveys for Jupiter-sized planets 
are inundated by astrophysical false positives, namely systems that contain an 
eclipsing binary and mimic the desired photometric signature.  We discuss 
several examples of such false alarms.  These systems were initially 
identified as candidates by the PSST instrument at Lowell Observatory.  
For three of the examples, we present follow-up spectroscopy that 
demonstrates that these systems consist of (1) an M-dwarf in eclipse 
in front of a larger star, (2) two main-sequence stars presenting grazing-incidence
eclipses, and (3) the blend of an eclipsing binary with the light
of a third, brighter star.  For an additional candidate, we present 
multi-color follow-up photometry during a subsequent time of eclipse, 
which reveals that this candidate consists of a blend of
an eclipsing binary and a physically unassociated star.
We discuss a couple indicators from publicly-available catalogs that 
can be used to identify which candidates are likely giant stars, a large 
source of the contaminants in such surveys.
\end{abstract}

\maketitle

%%%%%%%%%%%%%%%%%%%%%%%%%%%%%%%%%%%%%%%%%%%%
%% MAINMATTER
%%%%%%%%%%%%%%%%%%%%%%%%%%%%%%%%%%%%%%%%%%%%

\section{Wide-Field Transit Surveys}

\begin{figure}[t]
\includegraphics[width=.74\textwidth]{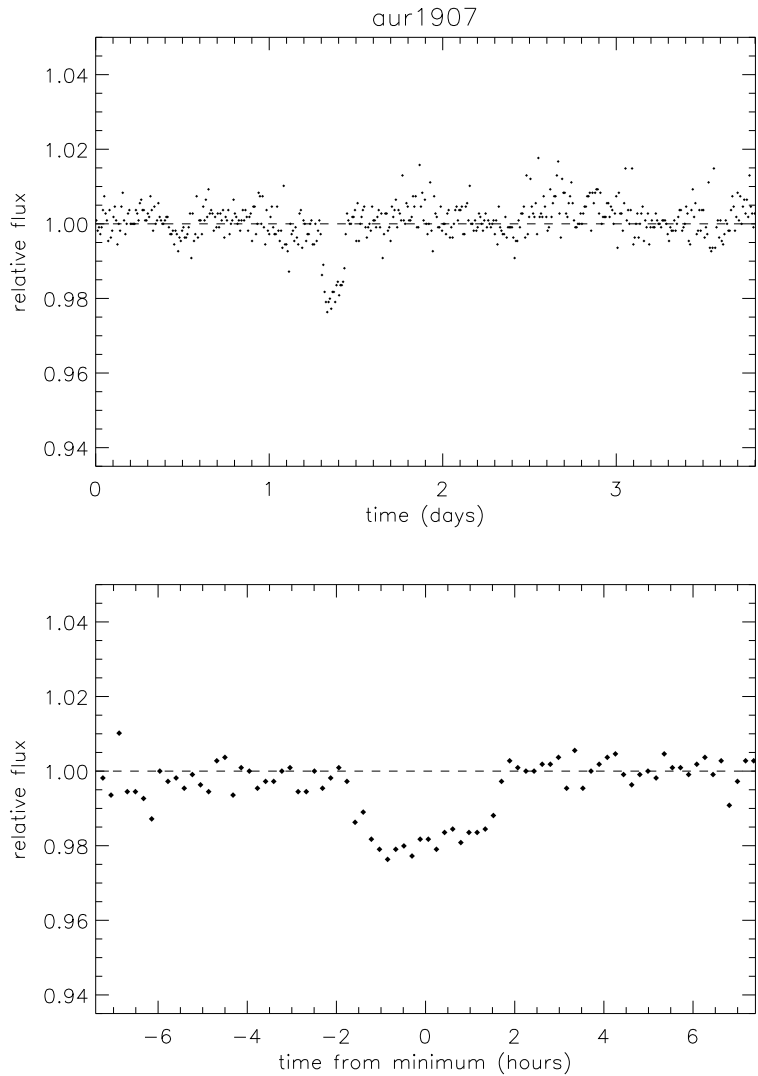}
\caption{(Upper panel) PSST photometric light curve for star~1907 ($V = 11.2$,
$B-V = 0.2$) in Auriga, binned and phased to a period of 3.80~days.  The light
curve is constant outside of eclipse.  (Lower panel) The same data
near times of transit.  The transit is flat-bottomed with a depth of 0.019~mag
and a duration of 3.7~hours, consistent with a transit of a Jovian planet across a 
Sun-like star.}
\end{figure}

\begin{figure}[t]
\includegraphics[width=.74\textwidth]{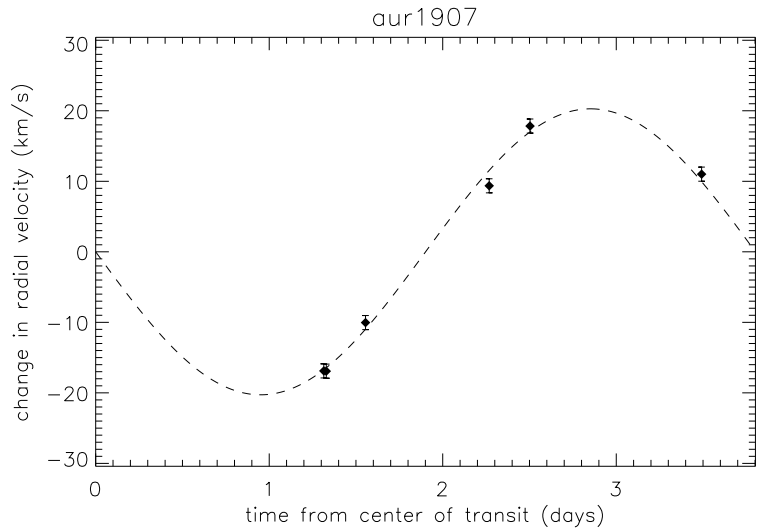}
\caption{Shown are the changes in radial velocity for star~1907 in Auriga (Figure~1)
as derived from high-resolution spectra gathered over 3 consecutive nights with the 
Palomar 60-inch echelle spectrograph.  The typical precision is $1\ {\rm km\, s^{-1}}$.
The observed variation is consistent with an orbit as predicted
by the photometric phase and period.  However, the amplitude of the
radial velocity orbit is $20\ {\rm km\, s^{-1}}$, indicating a 
low-mass stellar companion.}
\end{figure}

\begin{figure}[t]
\includegraphics[width=.74\textwidth]{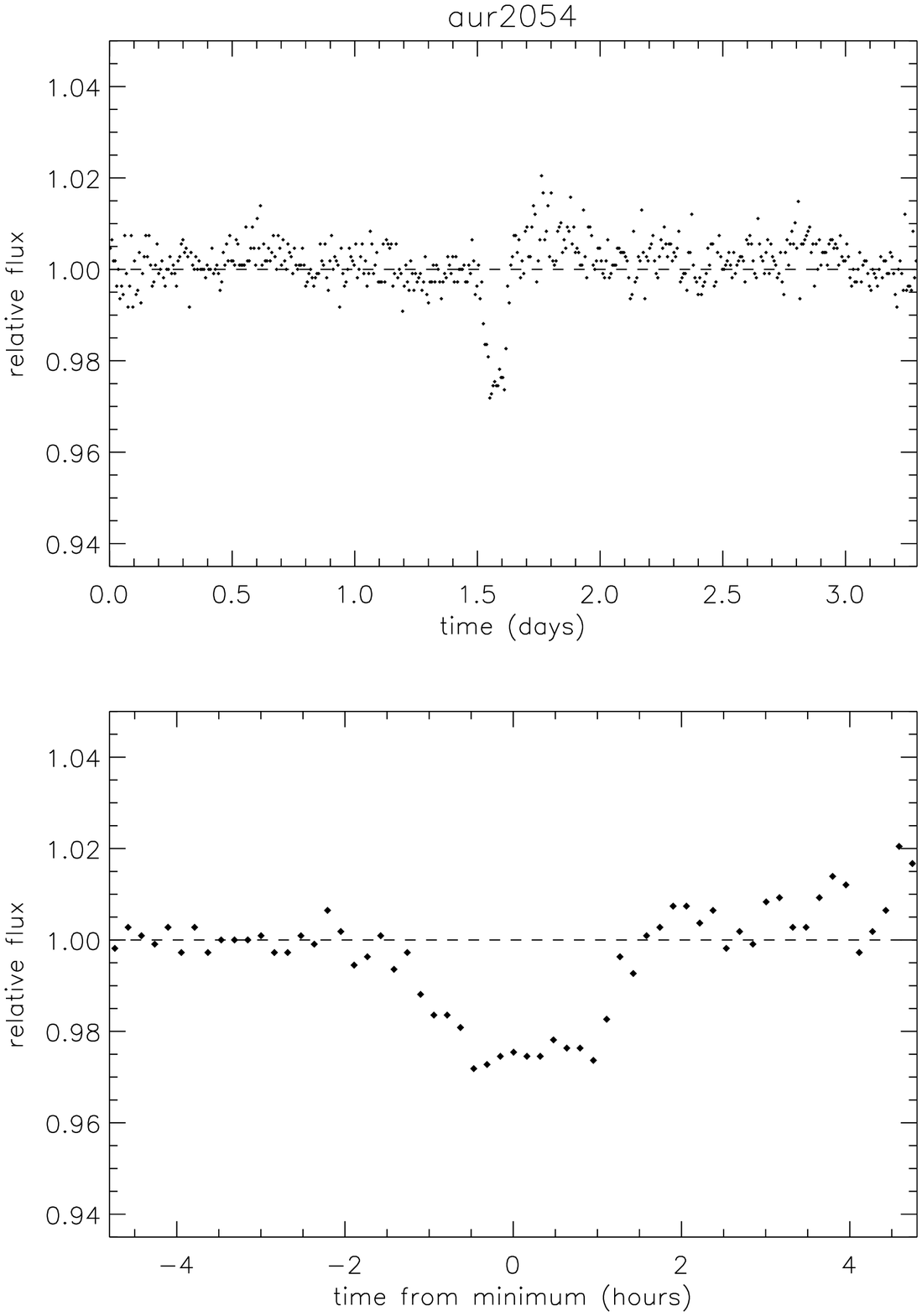}
\caption{(Upper panel) PSST photometric light curve for star~2054 ($V = 11.3$,
$B-V = 1.5$) in Auriga, binned and phased to a period of 3.29~days.  The light
curve is constant outside of eclipse.  (Lower panel) The same data
near times of transit.  The transit is flat-bottomed with a depth of 0.023~mag
and a duration of 2.4~hours, consistent with a transit of a Jovian planet across a 
Sun-like star.}
\end{figure}

\begin{figure}[t]
\includegraphics[width=.74\textwidth]{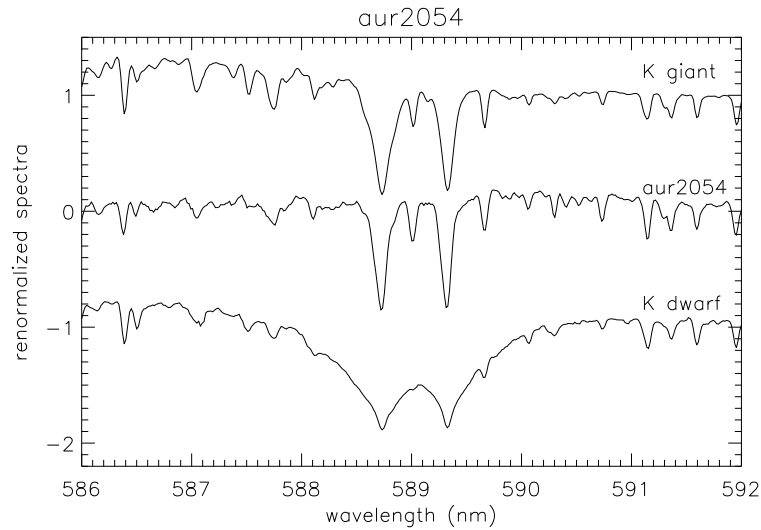}
\caption{The central curve shows a Palomar 60-inch echelle
spectrum of star~2054 in Auriga (Figure~3)
in the region of the gravity-sensitive sodium~D lines.  Spectra
of two K-stars with approximately the same color are also shown:
The lower spectrum is that of the K5-dwarf GJ~380, demonstrating the
very broad sodium lines indicative of the high gravity of
a dwarf star.  The upper spectrum is that of the K5.5-giant HR~5200,
where the sodium lines are unbroadened.  The semblance of the spectrum
of the candidate star to that of HR~5200 indicates that the
star is a giant.  Since a 3.3-day period would place the object within
the physical radius of the giant, this system is
likely a blend of K-giant and a fainter eclipsing binary.}
\end{figure}

\begin{figure}[t]
\includegraphics[width=.74\textwidth]{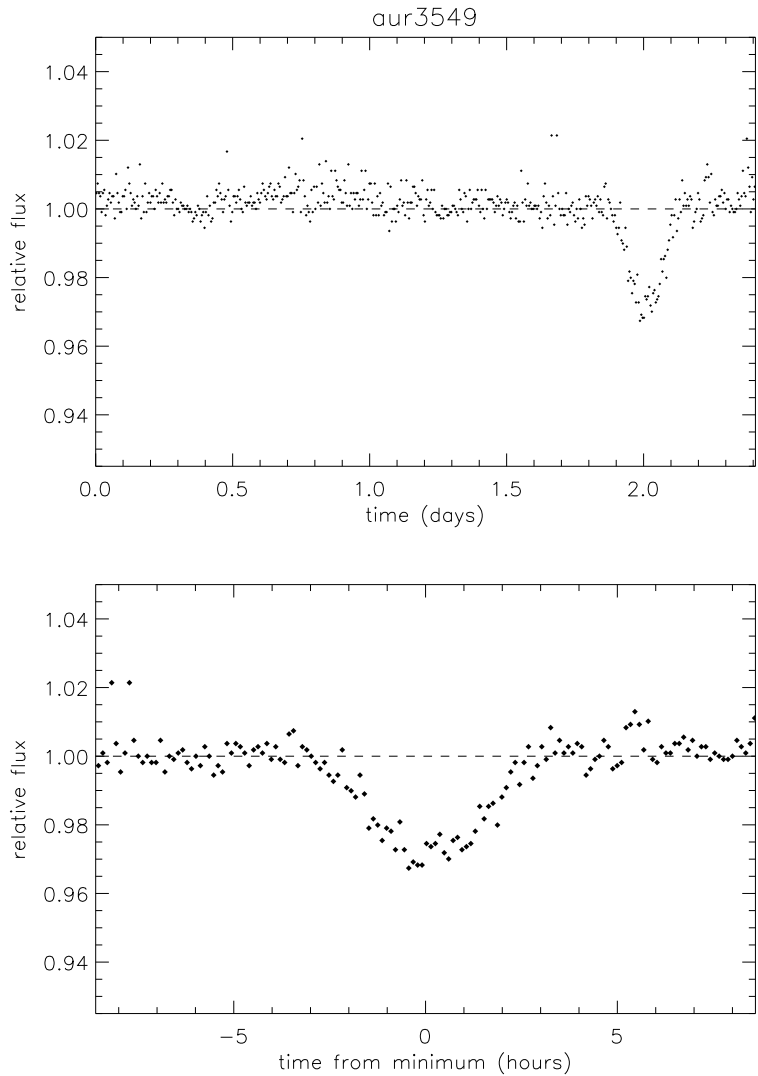}
\caption{(Upper panel) PSST photometric light curve for star~3549 ($V = 11.6$,
$B-V = 1.1$) in Auriga, binned and phased to a period of 2.41~days.  There
is some evidence for a secondary eclipse in the light curve, near a time
of 0.4~days.  (Lower panel) The same data near times of transit.  The transit 
is somewhat V-shaped, with a depth of 0.028~mag and a duration of 4.3~hours.
Although marginally consistent with a planetary transit, these data hint
at an eclipsing binary system.}
\end{figure}

\begin{figure}[t]
\includegraphics[width=.74\textwidth]{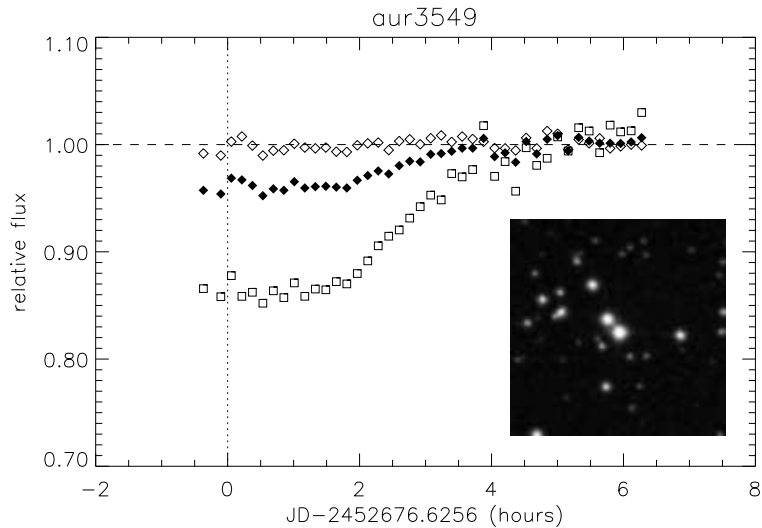}
\caption{The 2 arcmin $\times$ 2 arcmin Digitized Sky Survey image (inset) 
of star~3549 in Auriga (Figure~5) reveals two sources within the PSF of the PSST
instrument.  Photometry of the field during the time of a subsequent eclipse
(the predicted time of center of transit is shown as $t = 0$ in this figure)
shows the central brighter source to be constant in time (open diamonds).
The fainter star to the NE undergoes a deep (14\%) eclipse (open squares).
When the light from both stars is summed in a single photometric aperture,
the light curve shown with black diamonds results.  This photometric
time series reproduces that observed by the PSST (Figure~5).  Thus, this candidate is 
the blend of the fainter eclipsing binary to the NE, and the brighter
physically-unassociated central star.}
\end{figure}

\begin{figure}[t]
\includegraphics[width=.74\textwidth]{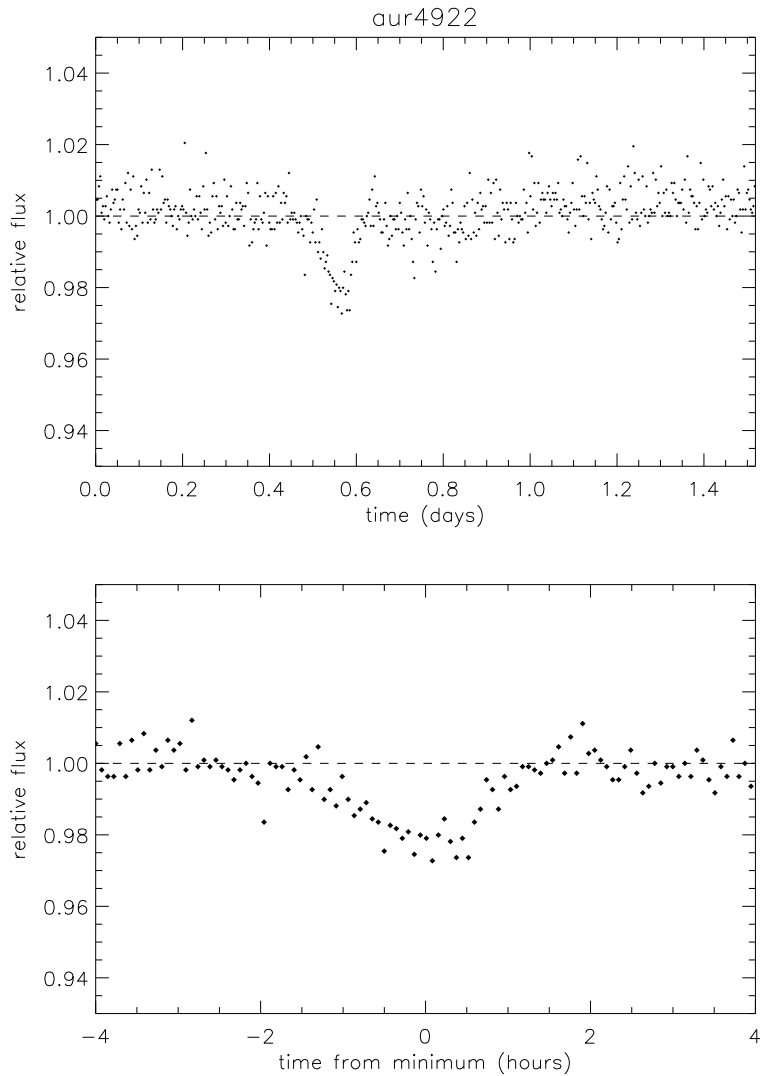}
\caption{(Upper panel) PSST photometric light curve for star~4922 ($V = 12.0$,
$B-V = 0.4$) in Auriga, binned and phased to a period of 1.52~days.  The
light curve is constant outside of eclipse.  (Lower panel) The same data near 
times of transit.  The transit is V-shaped, with a depth of 0.021~mag and a duration 
of 2.0~hours, indicating that the signal is likely due to a grazing-incidence
binary system.}
\end{figure}

\begin{figure}[t]
\includegraphics[width=.74\textwidth]{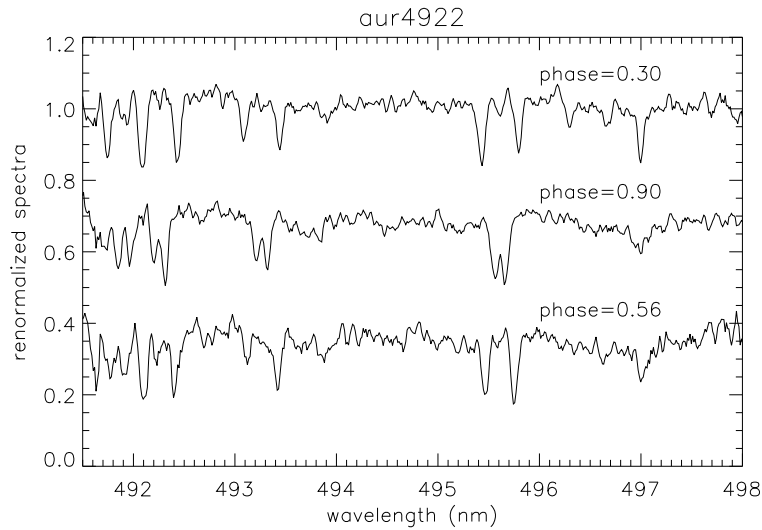}
\caption{Shown are Palomar 60-inch echelle spectra of star~4922 in Auriga (Figure~7) 
gathered on three consecutive nights.  The spectra are clearly double-lined.
The variations in radial velocity are consistent with the time scale of
the photometric period (the spectra are labeled by the orbital phase 
predicted by the photometry).  The radial velocity orbit is approximately
$90\ {\rm km\, s^{-1}}$.  Combined with the V-shaped nature of the eclipses,
we conclude that this is likely a grazing-incidence binary, with a true period 
that is twice the photometric period.}
\end{figure}

\begin{figure}[t]
\includegraphics[width=.74\textwidth]{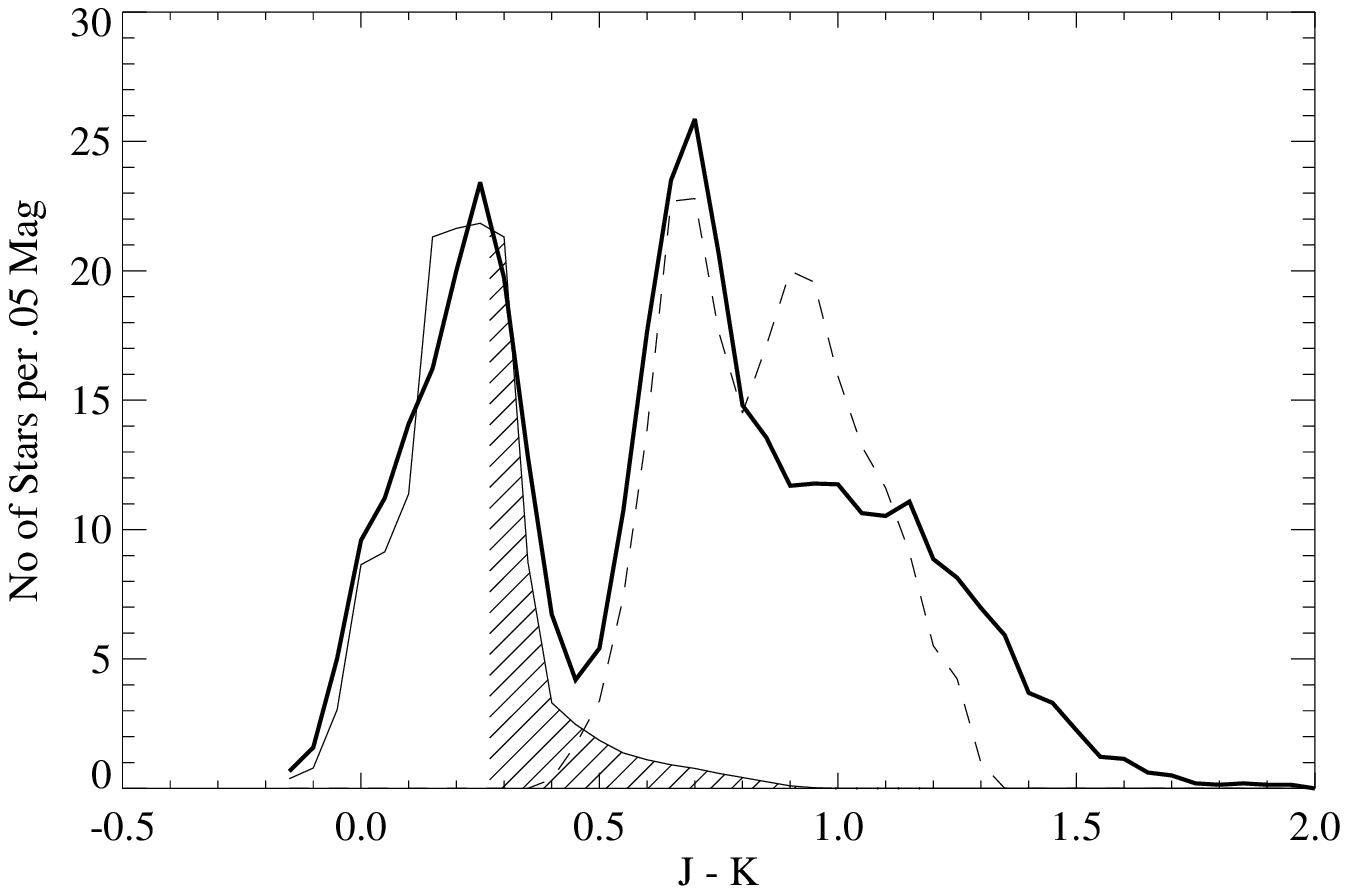}
\caption{Taken from \cite{brown2003}, courtesy T.~M.~Brown.  
The thick curve shows
the histogram of star density vs $J - K$ color from the 2MASS
catalog for stars observed by a STARE in a field in Cygnus.  The thin solid
curve shows the model used by Brown~\cite{brown2003} to predict
the occurence of main-sequence stars, and the dashed curve shows that for giants.  The
hatched region shows stars with radii less than 1.3~$R_{\odot}$.
Thus, the $J - K$ color appears to be a useful discriminant
against giant stars.}
\end{figure}

More than 20 groups worldwide are now engaged in photometric
surveys aimed at detecting Jupiter-sized planets in tight
orbits about their parent stars \cite[for a review, see ][]{horne2003, 
charbonneau2003b}.  Several of these projects are small, automated 
systems with modest apertures (typically 10~cm) and CCD cameras
(typically 2k~$\times$~2k arrays of $\sim15~\mu$m pixels), which 
monitor several thousand stars simultaneously in a very wide
field-of-view ($6\degr - 9\degr$~square).  Examples of such instruments
are STARE\footnote{\url{http://www.hao.ucar.edu/public/research/stare/stare.html}}
\cite[][located on Tenerife]{brown2000}, PSST~[located in northern Arizona], 
Sleuth\footnote{\url{http://www.astro.caltech.edu/~ftod/sleuth.html}} 
\cite[][located in southern California]{odonovan2003}, 
Vulcan\footnote{\url{http://web99.arc.nasa.gov/~vulcan/ }} 
\cite[][located in central California]{borucki2001, jenkins2002} and Vulcan 
South\footnote{\url{http://web99.arc.nasa.gov/~vulcan/south }} [to be located
at the South Pole], HAT\footnote{\url{http://cfa-www.harvard.edu/~gbakos/HAT/}} 
\cite[][located in southern Arizona]{bakos2002},
and SuperWASP\footnote{\url{http://www.superwasp.org/}} 
\cite[][located on La Palma]{street2003}.  The advantage that all such
projects offer over deeper transit surveys is the brightness 
of the target stars (typically $8 \le V \le 13$), which facilitates
radial velocity measurements aimed at detecting the orbit induced
by the planet.  Furthermore, it is only for such bright systems
that the host of follow-up measurements that are currently
being pursued for HD~209458 \cite{charbonneau2003a} and other bright
extrasolar-planet stars \cite{charbonneau2003c} might also be enabled.

The difficulty facing such surveys is not that of obtaining the
requisite photometric precision or phase coverage, as several
of the aforementioned projects have achieved these requirements.
Rather, the current challenge is that of efficiently rejecting
astrophysical false positives, i.e. systems containing an eclipsing binary,
whose resulting photometric light curves mimic that of a Jupiter
passing in front of a Sun-like star.

Brown \cite{brown2003} recently presented detailed estimates
of rates of such false alarms for these wide-field surveys.
The three dominant sources of such signals are (1) grazing incidence
eclipses in systems consisting of two main-sequence stars,
(2) central transits of low-mass stars in front of large
main-sequence stars (typically M-dwarfs passing in front of
F-stars), and (3) binary systems undergoing deep eclipses, blended with the
light of a third star that falls within the instrumental point-spread-function 
(the blending star may be either physically associated
or simply lying along the line-of-sight).  Brown finds that these
various forms of astrophsyical false positives will likely occur at
a combined rate nearly 12 times that of true planetary transits.  
A direct means to investigate the true nature of each candidate
would be high-precision radial-velocity monitoring, but this
would impose a hefty burden upon the large telescopes 
that are required.  Thus, all such transit surveys must adopt
strategies for efficiently rejecting the majority of such false alarms,
to bring the number of candidates to a manageable level.

\section{Examples of Astrophysical False Positives}
For the benefit of researchers pursuing such transit surveys,
we present here several examples of astrophysical false positives.
These candidate transit signals were identified in a survey of 
a $6\degr \times 6\degr$ field in Auriga by the PSST instrument 
at Lowell Observatory.  A description of the instrument and
the results of the survey will be presented elsewhere.
Our goal here is to present the follow-up
spectroscopy performed with the Palomar 60-inch telescope, 
which demonstrates conclusively that these candidates
are indeed false alarms.  Access to telescopes in the range 
of $1-2$~m is widely available.
As a result, such observations are a reasonable
means by which to dispose of such contaminants.  Similar
work to identify the true nature of candidates arising from
the Vulcan survey was presented in \cite{latham2003}.

\subsection{Star 1907:  An M-dwarf in eclipse}
The PSST photometric light curve (Figure~1) for star~1907 ($V = 11.2$,
$B-V = 0.2$) revealed a flat-bottomed transit with a period of 3.80~days,
a depth of 0.019~mag and a duration of 3.7~hours. The light
curve is constant outside of eclipse.  However, radial velocities (Figure~2)
derived from high-resolution spectra gathered over 3 consecutive nights with the 
Palomar 60-inch echelle spectrograph (with a typical precision of $1\ {\rm km\, s^{-1}}$)
revealed an orbit consistent with the period and phase derived by the
photometry, but with an amplitude of $20\ {\rm km\, s^{-1}}$.  This
large Doppler variation indicates a low-mass stellar companion.

\subsection{Star 2054: An eclisping binary blended with a giant star}
The photometric time series (Figure~3) for star~2054 ($V = 11.3$,
$B-V = 1.5$) shows a flat-bottomed eclipse with a period of 3.29~days,
a depth of 0.023~mag, and a duration of 2.4~hours.  The light
curve is constant outside of eclipse.  Figure~4 shows a
portion of the spectrum of this object gathered with the Palomar
60-inch echelle in the region of the gravity-sensitive sodium~D lines.  
The unbroadened profile of the sodium features reveals this star
to be a K-giant.  Since a 3.3~day orbit about a K-giant would place
it within the physical radius of the star, we must be seeing
a blend of a K-giant and a fainter eclipsing binary (which may
or may not be physically associated).

\subsection{Star 3549: An eclipsing binary blended with an unassociated star}
The photometric light curve (Figure~5) for star~3549 ($V = 11.6$,
$B-V = 1.1$) shows a mildly V-shaped eclipse with a period of 2.41~days,
a depth of 0.028~mag, and a duration of 4.3~hours.  Our suspicions
were raised by the hint of a faint secondary eclipse in the light curve.
Examination of the Digitized Sky Survey image (Figure~6) 
reveals two sources within the PSF of the PSST instrument.  
Photometry of the field during the time of a subsequent eclipse 
(conducted with a 14-inch telescope, which afforded a significant increase in the angular
resolution over the data gathered with the PSST)
finds that the central brighter source is constant, whereas
the fainter star to the NE undergoes a deep (14\%) eclipse (Figure~6).
When the light from both stars in summed in a single photometric aperture,
the light curve that results reproduces that observed by the PSST.
Hence, this system is a blend of an eclipsing binary (the fainter
star to the NE), and the physically unassociated brighter star.

\subsection{Star 4922:  A grazing-incidence eclipsing binary}
The photometric light curve (Figure~7) for star~4922 ($V = 12.0$,
$B-V = 0.4$) reveals a clearly V-shaped eclipse with a period of 
1.52~days, a depth of 0.021~mag, and a duration of 2.0~hours.  
The light curve is constant outside of eclipse, but the
shape of the transit already hints at a grazing-incidence
eclipsing binary.  Palomar 60-inch echelle spectra (Figure~8)
gathered on three consecutive nights clearly show two sets
of absorption features, and nightly variations in the radial velocity of the components 
that are consistent with the time scale of the photometric period.
The amplitude of the radial velocity orbit is approximately
$90\ {\rm km\, s^{-1}}$.  Combined with the V-shaped nature of the eclipses,
these data lead us to conclude that this is a grazing-incidence 
main-sequence binary, with a true period 
that is twice the photometric period.

\section{Efficient Methods to Reject False Alarms}
The publicly-available USNO-B 
catalog\footnote{\url{http://www.nofs.navy.mil/projects/pmm/catalogs.html}}
\cite{monet2003} provides proper motions, which aid in constraining
the distance of a star.  Typical errors in the proper motions from the USNO-B catalog
are $\sim7$~mas/yr.  Objects with colors consistent with
the target spectral types of K,G,or F and a high measured proper motion
are likely dwarf stars, and thus good candidates.  
Unfortunately, small proper motions are inconclusive:  
These stars could either be distant (hence giants), or nearby 
dwarfs.  In addition, colors from the 
2MASS catalog\footnote{\url{http://www.ipac.caltech.edu/2mass/releases/allsky/}} 
aid in separating dwarf stars from giants.  
Brown \cite{brown2003} illustrates that the $J-K$ colors from 
the 2MASS catalog provide a useful discriminant:  Stars with $J-K \ge 0.5$ are
predominantly giants, whereas those with $J-K \le 0.35$ are
almost exclusively dwarfs (Figure~9).  The target stars of wide-field transit
surveys are those with radii less than 1.3~$R_{\odot}$.  The
infrared colors of such stars extend into 
the intermediate region, so that the $J-K$ color is not a definitive test.
Nonetheless, candidates with very red colors ($J-K > 0.7$) are almost
certainly giants and can be rejected.

While the spectroscopy described in the previous section provides
a reliable means to identify astrophysical false positives, it is labor- 
and resource-intensive.  In order to more efficiently reject such contaminants,
L.~Kotredes and D.~Charbonneau are currently assembling Sherlock,
an automated follow-up telescope for
wide-field transit searches \cite{kotredes2003}.  Working from a list
of transit candidates, Sherlock will calculate future times of eclipse,
and conduct multi-color photometry of these targets, with an
angular resolution that is significantly improved over the survey instruments.  
The large majority of false alarms should yield eclipse signals that 
are color-dependent, whereas planetary transits should be nearly 
color-independent (except for the minor effects due to limb-darkening).  
Sherlock will be deployed at Palomar Observatory in early 2004, and will
be available to follow up candidates identified by any of the
wide-field transit searches.

\end{document}